# Social Aspects of Software Testing: Comparative Studies in Asia

Luiz F. Capretz[*,1][0000-0001-6966-2369], Jingdong Jia[2], Pradeep Waychal[3], Shuib Basri[4]

[1]Western University, London N6G5B9, Canada
lcapretz@uwo.ca
[2]Beihang University, Beijing, China
jiajingdong@buaa.edu.cn
[3]Chhatrapati Shadu Institute, Thane, India
Pradeep.waychal@gmail.com
[4]Universiti Teknologi PETRONAS, Malaysia
shuib_basri@utp.edu.my

**Abstract.** This study attempts to understand motivators and de-motivators that influence the decisions of software students to take up and sustain software testing careers across three different Asian countries, i.e., China, India, and Malaysia. The research question can be framed as "How many software students across different Asian geographies are keen to take up testing careers, and what are the reasons for their choices?" Towards an answer, we developed a cross-sectional but simple survey-based instrument. In this work, we investigated how software students perceived the software testing role. The results from China and India revealed that students are not very keen on taking up a software tester career, but the Malaysia students' show a more positive attitude towards software testing. The study also pointed out the importance of considering software testing activities as a set of human-dependent tasks and emphasized the need for further research that examines critically individual assessments of software testers about software testing activities. This investigation can academics involved in software testing courses to understand the impacting factors on the motivation and de-motivators of their students, as well as to try convey positive view of testing as challenging and requires critical thinking and innovative ideas.

**Keywords:** Social Aspects of Software Testing, Human Factors in Software Engineering, Software Testing, Software Careers, Cross-Cultural Study.

## 1     Introduction

The importance of software quality has been on the rise and is expected to grow in the future. Software testing plays a vital role in the development of high quality and reliable software systems. Software systems are becoming more complex due to the intricacies of the problems that they are solving, their dependence on a myriad of third-party software components, and the continuous shrinking of time to deployment. That complexity is inducing increased susceptibility to failures. Coupled with that, the pervasiveness of software systems, especially in mission-critical areas, is resulting in dangerous combinations. Vigilant testing with the help of appropriate processes, technology, and professionals can ease the situation. Therefore, the importance of software testing has been on the growth trajectory.

In spite of its relevance, software testing is, arguably, the least understood part of the software life cycle and still the toughest to perform correctly. Many researchers and practitioners have been working to address the situation [1]. However, most of the studies focus on the process and technology dimensions [2] and only a few on the human dimension of testing [3], despite the reported relevance of human aspects during testing [4]. Testers need to understand various stakeholders' explicit and implicit requirements, be aware of how developers work individually and in teams, and develop skills to report test results wisely to stakeholders. These multifaceted requirements lend vitality to the human dimension in software testing. Exploring this dimension carefully may help understand testing in a better way.

This study attempts to solve the basic problem of the human dimension of testing, i.e. the lack of competent testing professionals, by trying to understand the unwillingness of software or computer engineering students across different geographies, and their reasons for not taking up testing careers. This paper compares samples from the India, China, and Malaysia – three Asian countries. Given an overall Asian culture, this study seeks to investigate any difference among Asian students in regard to software testing. The research question, therefore, is; why software and computer engineering students (henceforth referred to as students) across different parts of the world are reluctant to consider software testing careers?

After this introduction, the rest of the paper is organized as follows. Section 2 describes the research method, instruments and techniques applied to answer our research questions, and section 3 presents our findings, which are discussed in Section 4. The implications of the study are given in section 5, and concluding remarks in section 6.

## 2  Method

Our study analyzed the opinions of students whether they would choose testing careers and what they felt were the advantages and drawbacks of the testing career.

### 2.1  Objectives

Software testing requires competent and motivated professionals [5]. However, very few students, across the globe, prefer testing careers, which robs the industry of competent testers and consequently, quality software. This study attempts to understand reasons for such apathy towards testing careers. The study is cross-sectional and mixed. It does not study students' responses over time and seeks open-ended responses to advantages and drawbacks of the testing career and categorical responses about choosing testing careers.

### 2.2 Instrument Selection

The instrument was designed to understand the willingness of students to take up software testing careers and the reasons thereof. Specifically, the students were asked for the probability of their choosing testing careers by selecting one of the following choices: 'Certainly Yes,' 'Yes,' 'Maybe,' 'No,' and 'Certainly Not.' The study asked the respondents to provide open-ended but prioritized list of advantages and drawbacks, and open-ended rationale regarding their decisions on taking up testing careers. Since there has been limited prior research in the area, especially in the geographies that we were studying, we decided to use such a qualitative approach to investigate and understand the phenomena within their real-life context.

Some researchers have distilled motivators and de-motivators of software profession [6], which could have formed a basis for our investigation. However, the three studies neither converged indicating a possibility of some more factors nor covered students' perceptions.

A fresh and simple survey questionnaire, therefore, was designed in consultation with researchers and testing professionals, which is presented in Appendix A. The authors carried out a pilot survey using a sample of 20 Indian students and ascertained that the students neither had any queries nor reported any lack of clarity in the questionnaire. The survey was translated to Chinese and the Chinese responses were translated back in English to check the accuracy of the Chinese questionnaire.

### 2.3 Sampling

The study used convenience sampling by seeking responses from 251 students from three different countries (70 from India, 99 from China, and 82 from Malaysia). The Indian responses were sought from junior students towards the end of their second semester, the Chinese responses from senior students before the start of their first semester, and the Malaysian responses from senior students towards the end of their first semester. While the Indians and Chinese students had not taken a course on software testing, the Malaysian students did. Thus students' curricular background, relevant for this study, and levels in the programs are comparable and their career decisions are likely to be due to the local situations and the essential nature of testing activities.

### 2.4 Reliability and Validity

A reliable and valid study indicates absence of bias and a high degree of truthfulness. A qualitative study needs to be trustworthy and rigorous, and of good quality. We asked the respondents to list advantages and drawbacks of testing careers and the probability that they would choose a testing career, along with their rationales. We triangulated the rationales and advantages/drawbacks and practically did not find any divergence between them. All the responses were coded by the authors of the respective countries and reviewed by all authors.

## 3 Data Collection: Results and Findings

The objectives of the study were explained to students and their responses to the survey were sought. They were assured that their responses would not influence course grades in any way and were offered an option of not disclosing their identities. The responses were manually tagged and iteratively coded until no further code changes were possible. The next subsection presents the probabilities of students taking up testing careers (Table 1), and the following subsections list advantages and drawbacks indicated by all respondents.

Table 1. Chances of students taking up testing careers.

| Response | India | China | Malaysia |
|---|---|---|---|
| Certainly Not | 14% | 24% | 1% |
| No | 31% | 0% | 7% |
| May be | 47% | 74% | 52% |
| Yes | 7% | 2% | 34% |
| Certainly Yes | 0% | 0% | 6% |

**China.** Most of the students (74%) chose the 'May Be' option. 24% were very clear that they would not take up the testing career and chose 'Certainly Not' and only 2% of students chose the 'Yes' option.

**India.** No student chose the 'Certainly Yes' option and only 7% of students chose the 'Yes' option. 14% of students selected 'Certainly Not' and 31% opted for the 'No' option. 47% percent of students were unsure and chose 'May Be'.

**Malaysia.** 40% of students were willing to be testers and 6% of them responded with 'Certainly Yes'. That's a huge number compared to the other two countries. Only 8% of the students were not ready to take up testing careers, and a mere 1% of them responded with 'Certainly No'. A significant 52% chose the 'May Be' option.

### 3.1 Advantages of Testing Careers

The analysis of responses to the advantages resulted in the following factors:

- Learning opportunities – Testers can learn about different products, technologies, techniques, and languages as well as domains such as retail, financial. They also can develop softer skills due to more interactions, many of them being difficult ones, with developers and customers. Testing activities provide the full background of a project's scope and architecture in a short period of time and may span all project stages. Further, testing requires focusing on details and is a growing field.
- Important job – Testers are accountable and responsible for the product quality. In that sense, testing is an important part of the software life cycle.

- Easy job – This refers to students' belief that testing does have well defined and easy processes, etc.
- Thinking job – This encompasses views about testing such as being challenging, creative, innovative, and requiring logical and analytical thinking.
- More job – This states that more testing jobs are available and due to the higher demands, the jobs are secure and stable.
- Monetary benefits – Testing jobs does not have bad salary packages.
- Fun to break things – Some students felt that it is fun to break things and find defects in software.

The responses from each country are analyzed and presented in Table 2. Since we excluded advantages that were less than 5% (too small to consider) of the total advantages, the total in each column may not be 100%.

Table 2. Percentages of Salient Advantages by country

| Advantages | India | China | Malaysia |
| --- | --- | --- | --- |
| Learning Opportunities | 30% | 8% | 53% |
| Important Job | 14% |  | 37% |
| Easy Job | 9% | 44% | 28% |
| Thinking job | 38% |  | 20% |
| More jobs |  | 22% | 37% |
| Monetary benefits |  | 13% | 29% |
| Fun to break things |  | 6% | 7% |

**China.** 44% and 22% of advantages referred to testing being easy and offering more jobs, respectively. 13% of advantages indicated to proper monetary rewards for testing professionals. 'Learning opportunities' (8%), and 'fun to break things' (6%) were the next set of advantages.

**India.** Indian students' most voted advantage was testing being a 'thinking job' (38%). Its learning opportunities and importance fetched 30% and 14% of advantages, respectively. Easiness of the jobs polled just 9% of advantages.

**Malaysia.** 53% of advantages referred to the learning opportunities the testing careers offered. 37% of advantages recognized the testing jobs to be important and the same percentage realized that there are more testing jobs available. While 29% asserted testing as having proper monetary incentives. Testing jobs were also seen as 'thinking job' (20%) and with the prospect of 'fun to break thing's' (7%).

### 3.2 Drawbacks of Testing Careers

The analysis of responses of the drawbacks resulted in the following factors:

- Second-class citizen – This points out that testers are not involved in decision making and are blamed for poor quality but developers are rewarded for good quality. They also include the evident lack of support from management in terms of unrealistic schedules, poor allocation of resources, and inadequate recognition.
- Career development – Students believe that there is limited growth in the testing field. Some also believe that testers' jobs are less secure and that they are the first ones to lose their jobs during business downturns.
- Complexity – This covers tester facing complex situations such as different versions of software products, platform incompatibilities, defects not getting reproduced, and testing not given sufficient time, but being held responsible for product quality. This also includes the fact that testers need to look at business and technology artifacts and understand many abstractions. The lack of clarity around requirements also adds to the difficulties.
- Tedious – This refers to the repetitive nature of testing and respondents have also used words such as monotonous and boring.
- Missed development – Some believe that they miss opportunities for professional development by taking up testing careers. While some testers do develop test automation systems, the students seem to consider that as different from actual development activity. Some also think that they lose learning opportunities that are available to developers.
- Less monetary benefits – Some testers believe that testers' jobs do not have monetary benefits at par with developers.
- Finding others' mistakes – It is not pleasant to find mistakes in others' work and tell them, although testers have to report anything that adversely impacts the value of the product, which may not go well with some stakeholders.
- Stressful job: This has to do with work-life balance.
- No Interest – Some students just mentioned that they have no interest in the field of software testing.

The responses of students from each country are analyzed and presented in Table 3. Since we excluded drawbacks that were less than 5% of the total drawbacks, the total in each column may not be 100%.

**China.** The Chinese students' highest votes went to complexity (37%) and tediousness (35%) of testing jobs. 'Missing development' fetched 7%, and 'limited career development' and 'less monetary benefits' polled around 6% each.

**India.** The Indian students' most important drawbacks were treatment as 'second-class citizens' (25%) and testing being complex (24%) and tedious (24%). 15% reported to 'missing development' and 5% to a lack of personal interest.

**Malaysia.** The most voted drawbacks for Malaysian students were tediousness (39%), missing development (18%) and treatment as second-class citizens (only 7%). Com-plexity and finding mistakes of others polled 33% and 34% each respectively, which are quite high percentages.

**Table 3.** Percentages of salient drawbacks by country

| Drawbacks | China | India | Malaysia |
|---|---|---|---|
| Second-class citizen | | 25% | 7% |
| Career development | 6% | | 29% |
| Complexity | 37% | 24% | 33% |
| Tedious | 35% | 24% | 39% |
| Missing development | 7% | 15% | 18% |
| Less monetary benefits | 6% | | 6% |
| Finding others' mistakes | | | 34% |
| Stressful job | | | 15% |
| No interest | | 5% | 17% |

## 4 Discussions

In this section, we gleaned through the main findings of our research. Testing jobs remain unpopular across China and India, but not in Malaysia which has a significant fraction of students (40%), who wanted to take up testing careers. It is, perhaps, due to having a prior exposure to testing through a course. This survey was taken by Malaysian students towards the end their course in software testing..

A large segment of the Chinese students was undecided as almost half of Chinese undergraduates take up graduate studies and not jobs. Many Indian students were against taking up testing careers, which may be due to the low unemployment rate. The higher percentage of naysayers from India and China may be because they not having a course on testing. In case of Malaysia, the naysayers were minimum, a rare 1% of the sample. This indicates the importance of proper treatment of testing in curricula, like the Malaysian curriculum.

Testing offers learning opportunities as reported by students from all the three countries, but not as the most voted advantage. That was, in the case of Indian students', testing is perceived as a 'thinking job'. Chinese students' most voted advantages were ease and the number of testing jobs and that of Malaysian students' were learning opportunities, importance and the number of jobs.

The most prominent drawbacks appeared to be tediousness in the case of Malaysian students, and both tediousness and complexity resulting in stress and frustration in the case of Chinese students. For Indian students, 'second-class citizen treatment' meted out to the testers, tediousness and the complexity were the most prominent drawbacks. The respondents from all the three countries indicated 'missing development' as a drawback with Chinese students not feeling so strongly about that.

There appeared to be some confusion among students as the list included contradictory factors such as easy and thinking jobs, less and more monetary benefits, and tedious and creative jobs. Moreover, some students expressed frustration arising from debugging, as if the testers would be correcting the software, a common misconception about testing and debugging.

## 5    Implications to Computer Science and Software Engineering Curricula

The study has many implications for colleges, especially for computer and software engineering departments, and for industry leaders.

Astigarraga et al. [7] present a survey of the software testing curriculum in the United States and discuss efforts made to improve the status of testing in the academic curriculum. They report minimal content on software testing in the undergraduate curriculum. Either the testing topics are included in courses such as software engineering, combined with quality assurance, or through independent but optional courses. The IEEE guide to software engineering body of knowledge SWEBOK [8] and ACM – IEEE Curriculum Guidelines for Undergraduate Degree Programs in Software Engineering [9] are the most widely used sources for developing curricula in software engineering and allied disciplines. Their treatment of software testing is limited to process and technology dimensions and does not bring out the thinking and communication challenges faced by software test engineers. Therefore, curricular changes may be required to bring out the right perspectives of testing activities.

Since testing courses can improve the perception of testing careers, colleges can introduce them in their curricula. They can regularly review the curricula by consulting their alumni and to ongoing research. Since testing offers additional jobs, the course can help colleges improve placement prospects of their students.

The testing curriculum needs to reflect the understanding that testers need to provide correct information to various stakeholders, and appreciate that testing is 'applied epistemology' grounded in 'cognitive psychology'. The faculty must dispel beliefs such as, testing is just mechanically running tests and comparing outputs with expected results. Instead, they should explain the importance of testing and the philosophy behind it and impress upon the students that any testing assignment and design of effective test cases can be very creative. They should develop testers who can understand different domains and the needs of users in those domains, to understand the developer mindset and anticipate mistakes that developers may be making as an individual and as a team, to test creatively and efficiently under the given constraints, and to report the findings wisely to all stakeholders.

Colleges can also create better awareness of the importance of testing by introducing deployable projects. The projects would help students understand the connection between software quality and testing. It would also give them a first-hand experience of the advantages of testing such as relevance and the thinking nature and learning opportunities including learning the critical soft skills.

Additionally, the advantages and drawbacks of testing careers, as perceived by students, can help test managers and team leaders scale the challenge of recruiting test professionals. Understanding the common as well as country-specific advantages and drawbacks may help managers dealing with global teams. As emphasized before, software testing is a human activity and testers, who willingly take up testing careers, can influence the quality of the final product.

Since 'tediousness' is coming out as a major drawback, industry leaders may want to introduce sufficient automation or recruit high-school graduates for testing techni-cian roles, and provide more creative work to engineering graduates. Automation also brings in the development element, which may take care of the drawback 'miss development'. Another important thing that industry leaders need to work on is eliminating the feeling of testers of being 'second-class citizens'; although students have not indicated this as a critical issue, professionals have done so [10], understandably because they are working on the environment. If students come to know about this factor, they will further veer away from the testing profession. Colleges can counter the 'second-class citizen' problem by organizing interactions with thought leaders in software testing.

## 6    Conclusions and Future Directions

This investigation presents an analysis of the survey responses of 251 computer and software engineering students from three different countries studying at similar levels, about their willingness to take up software testing careers and the factors that influence their decisions.

The general empirical findings on the advantages as perceived by students from these countries do not seem to converge. While learning opportunities and easiness of the job are the common advantages, their percentages vary widely from 8% to 44%. While Indian students' major advantages are 'testing is a thinking job' and 'testing offers learning opportunities', for Chinese students it is 'easiness and plenty of testing jobs', and for Malaysian are the 'learning opportunities, importance and the large number of testing jobs'. In case of drawbacks, there is relatively a better convergence. The common drawbacks are tediousness, complexity, and 'missing development', although the range varies widely from 6% to 50%. 'Treatment as second-class citizen' figures high in the drawbacks Indian students, but it is irrelevant for Malaysian students.

It is heartening to see that there are more work-related advantages, such as, 'challenging job' and 'important job' that relate to intrinsic motivation than environmental/hygiene factors such as 'monetary rewards'. However, it is important that students feel that they will be valued and respected and not treated as 'second-class citizens' while working as testers. Since this study provides factors that were identified by many survey respondents, they may be transferable to other situations.

The study has limitations with respect to the sample studied. As a future direction, increasing the number of participants with more variation of gender, curricula of students, and geographies (covering European and Pacific Rim nations) is required. Further, willingness to take up a career is influenced by industry dynamics as well as colleges' dynamics, which may require a longitudinal study to better understand the phenomenon. Despite these limitations, the study offers insights useful for educational institutions to appropriately develop their students, and industry practitioners to tackle recruitment and retention problems of software testing professionals.

355

# Appendix A – Survey Questions

1. What are the three PROs (in the order of importance) for taking up testing career?
      a)                 b)                 c)
2. What are three CONs (in the order of importance) for taking up testing career?
      a)                 b)                 c)
3. What are chances of my taking up testing career?
          Certainly Not   No   Maybe   Yes   Certainly Yes
          Reasons:
4. Gender (optional):
5. GPA (optional):